\documentclass[twocolumn,floatfix,amssymb,amssymb,aps,pra]{revtex4} 
\usepackage{epsfig}

\begin{document}

\title{Stability of general relativistic static thick disks: the
isotropic Schwarzschild thick disk}

\author{Maximiliano Ujevic}\email[e-mail: ]{mujevic@ime.unicamp.br}
\author{Patricio S. Letelier}\email[e-mail: ]{letelier@ime.unicamp.br}
\affiliation{Departamento de Matem\'atica Aplicada, Instituto de
Matem\'atica, Estat\'{\i}stica e Computa\c{c}\~ao Cient\'{\i}fica \\
Universidade Estadual de Campinas, 13081-970, Campinas, S\~ao Paulo,
Brasil}

\begin{abstract}

We study the stability of general relativistic static thick disks, as
an application we consider the thick disk generated by applying the
``displace, cut, fill and reflect" method, usually known as the image
method, to the Schwarzschild metric in isotropic coordinates. The
isotropic Schwarzschild thick disk obtained from this method is the
simplest model to describe, in the context of General Relativity, real
thick galaxies. The stability under a general first order perturbation
of the disk is investigated. The first order perturbation, when
applying to the conservation equations, leads to a set of differential
equations that are, in general, not self-consistent. This opens the
possibility of performing various kinds of perturbations to transform
the resulting system of equations into a self-consistent system. We
perform a complete classification of the perturbations as well as the
stability analysis for all the relevant physical perturbations. We
found that, in general, the isotropic Schwarzschild thick disk is
stable under these kinds of perturbations.

\end{abstract}

\maketitle

\section{Introduction} 

Exact axial symmetric solutions of Einstein field equations are a good
starting point for modeling astrophysical applications in General
Relativity. This is due to the fact that the natural shape of an
isolated self-gravitating fluid is axially symmetric. In the last
decades, several exact solutions were studied as possible galactic
models. Static thin disk solutions were first studied by Bonnor and
Sackfield \cite{bon:sac} and Morgan and Morgan \cite{mor:mor1}, where
they considered disks without radial pressure. Disks with radial
pressure and with radial tension have been considered by Morgan and
Morgan \cite{mor:mor2} and Gonz\'alez and Letelier \cite{gon:let1},
respectively. Self-similar static disks were studied by Lynden-Bell and
Pineault \cite{lyn:pin}, and Lemos \cite{lem}. Moreover, solutions that
involve superpositions of black holes with static disks were analyzed
by Lemos and Letelier \cite{lem:let1,lem:let2,lem:let3} and Klein
\cite{kle1}. Also, relativistic counter-rotating thin disks as sources
of the Kerr type metrics were found by Bi\v{c}\'ak and Ledvinka
\cite{bic:led}. Counter-rotating models with radial pressure and dust
disks without radial pressure were studied by Gonz\'alez and Espitia
\cite{gon:esp}, and Garc\'{\i}a and Gonz\'alez \cite{gar:gon},
respectively; while rotating disks with heat flow were studied by
Gonz\'alez and Letelier \cite{gon:let2}. Furthermore, static thin disks
as sources of known vacuum spacetimes from the Chazy-Curzon metric
\cite{cha,cur} and Zipoy-Voorhees \cite{zip,voo} metric were obtained
by Bi\v{c}\'ak, Lynden-Bell and Katz \cite{bic:lyn1}. Also,
Bi\v{c}\'ak, Lynden-Bell and Pichon \cite{bic:lyn2} found an infinite
number of new static solutions. Stationary disk models including
electric fields \cite{led:zof}, magnetic fields \cite{let}, and both
electric and magnetic fields \cite{kat:bic} have been studied. In the
last years, exact solutions for thin disks made with single and
composite halos of matter \cite{vog:let1}, charged dust \cite{vog:let2}
and charged perfect fluid \cite{vog:let3} were obtained. For a survey
on relativistic gravitating disks, see \cite{sem,kar:hur}. Most of the
models constructed above were found using the metric to calculate its
energy momentum-tensor, i.e. a inverse problem. Several exact disk
solutions were found using the direct method that consists in computing
the metric for a given energy momentum tensor representing the disk
\cite{neu:mei, kle:ric, kle2, fra:kle, kle3, kle4, kle5}. In a first
approximation, the galaxies can be thought to be thin, this usually
simplifies the analysis and provides very useful information. But, in
order to model real physical galaxies the thickness of the disks must
be considered. Exact axially symmetric relativistic thick disks in
different coordinate systems were studied by Gonz\'alez and Letelier
\cite{gon:let3}. Also, different thick disks were obtained from the
Schwarzschild metric in different coordinates systems with the
``displace, cut, fill, and reflect" method \cite{vog:let4}. Recently, a
relativistic generalization of the Miyamoto-Nagai potential was
obtained \cite{vog:let5}.

The applicability of these disks models to any structure found in
Nature lays in its stability. The study of the stability, analytically
or numerically, is vital to the acceptance of the particular model.
Also, the study of different types of perturbations, when applied to
these models, might give an insight on the formation of bars, rings or
different stellar patterns. Furthermore, a perturbation can cause the
collapse of a stable object with the posterior appearance of a
different kind of structure. An analytical treatment of the stability
of disks in Newtonian theory can be found in \cite{bin:tre,fri:pol} and
references therein. In general, the stability of disks in General
Relativity is done in two ways. One way is to study the stability of
the particle orbits along geodesics. This kind of studied was made by
Letelier \cite{let2} transforming the Rayleigh criterion of stability
\cite{ray,lan:lif} into a general relativistic formulation. Using this
criterion, the stability of orbits around black holes surrounded by
disks, rings and multipolar fields were analyzed \cite{let2}. Also,
this criterion was employed in \cite{vog:let1} to study the stability
of the isotropic Schwarzschild thin disk, and thin disks of single and
composite halos. The stability of circular orbits in stationary
axisymmetric spacetimes are studied in \cite{bar,abr:pra}. Moreover,
the stability of circular orbits of the Lemos-Letelier solution
\cite{lem:let2} for the superposition of a black hole and a flat ring
are considered in \cite{sem:zac, sem:zac2, sem2}. Also, Bi\v{c}\'ak,
Lynden-Bell and Katz \cite{bic:lyn1} analyzed the stability of several
thin disks without radial pressure or tension studying their velocity
curves and specific angular momentum. The other way to study the
stability of disks is to perturb its energy momentum tensor. This way
is more complete than the analysis of particle motions along geodesics
because we are taking into account the collective behavior of the
particles. However, there are few studies in the literature performing
this kind of perturbation. A general stability study of a relativistic
fluid, with both bulk and dynamical viscosity, was done by Seguin
\cite{seg}. He considered the coefficients of the perturbed variables
as constants, i.e. local perturbations. Usually, this condition is too
restrictive. Stability analysis of thin disks from the Schwarzschild
metric, the Chazy-Curzon metric and Zipoy-Voorhees metric perturbing
their energy momentum tensor with a general first order perturbation
were made by Ujevic and Letelier \cite{uje:let}, finding that the thin
disks without radial pressure are not stable \cite{uje:let2}.

In order to have a general relativistic physical model for galaxies, we
must considered, first of all, the thickness of the disk and its
stability under perturbations of the fluid quantities. The purpose of
this work is to study numerically the stability of the isotropic
Schwarzschild thick disk under a general first order perturbation. The
perturbation is done in the temporal, radial, axial and azimuthal
components of the quantities involved in the energy momentum tensor of
the fluid. In the general thick disk case, the perturbed conservation
equations are not self-consistent because the number of unknowns is
larger than the number of equations. In order to make the system of
equation self-consistent we must chose a combination of the perturbed
quantities. This opens the possibility to perform several types of
perturbed quantities combinations that provide a self-consistent set of
equations. In this manuscript, we study the physical accepted
perturbations. We mean by a physical perturbation the fact that a
perturbation in a given direction of the pressure must create a
perturbation in the same direction of the four velocity. The energy
momentum perturbation considered in this manuscript is treated as
``test matter", so it does not modified the background metric obtained
from the solution of Einstein equations. 

The article is organized as follows. In Sec. II, we present the general
perturbed conservation equations for the thick disk case. The energy
momentum tensor is considered diagonal with its elements different from
zero. Also, the energy momentum perturbation is considered as ``test
matter". In this section we classified the possible perturbed
quantities combinations that leads to a self-consistent set of the
perturbed conservation equations of motion. In Sec. III, we present the
thick disk considered for the analysis, i.e. the isotropic
Schwarzschild thick disk. The form of its energy density and pressures,
as well as, the restrictions that the thermodynamic quantities must
obey to satisfy the strong, weak and dominant energy conditions are
shown. Later, in Sec. IV, we perform all the physical accepted
perturbations to the isotropic Schwarzschild thick disk to study its
stability. Finally, in Sec. V, we summarized our results.

\section{Perturbed Equations}

The thick disk considered is a particular case of the general static,
axially symmetric metric 

\begin{equation}
ds^2 = -e^{2 \Psi_1} dt^2 + e^{2 \Psi_2} R^2 d\theta^2 + e^{2 \Psi_3}
(dR^2 + dz^2), \label{metric}
\end{equation}

\noindent where $\Psi_1$, $\Psi_2$ and $\Psi_3$ are functions of the
variables ($R,z$). (Our conventions are: $G=c=1$. Metric signature +2.
Partial and covariant derivatives with respect to the coordinate
$x^\mu$ denoted by $,\mu$ and $;\mu$, respectively.)

In its rest frame, the energy momentum tensor of the fluid $T^{\mu\nu}$
is diagonal with components (-$\rho,p_R,p_\theta,p_z$), where $\rho$ is
the energy density and ($p_R,p_\theta,p_z$) are the radial, azimuthal
and axial pressures or tensions, respectively. So, in this frame of
reference, the energy momentum tensor can be written as

\begin{equation} 
T^{\mu\nu} = \rho U^\mu U^\nu + p_R X^\mu X^\nu + p_\theta Y^\mu Y^\nu
+ p_z Z^\mu Z^\nu, \label{tmunu} 
\end{equation}

\noindent where $U^\mu$, $X^\mu$, $Y^\mu$, and $Z^\mu$ are the four
vectors of the orthonormal tetrad

\begin{eqnarray}
&&U^\mu = e^{-\Psi_1} (1,0,0,0), \nonumber \\
&&X^\mu = e^{-\Psi_3} (0,1,0,0), \nonumber \\
&&Y^\mu = \frac{e^{-\Psi_2}}{R} (0,0,1,0), \nonumber \\
&&X^\mu = e^{-\Psi_3} (0,0,0,1), \label{tetrad}
\end{eqnarray}

\noindent which satisfy the orthonormal relations. Note that with the
above definitions, the timelike four velocity of the fluid is $U^\mu$
and the quantities $X^\mu$, $Y^\mu$, and $Z^\mu$ are the spacelike
principal directions of the fluid. Furthermore, the energy momentum
tensor satisfies Einstein field equations, $G_{\mu\nu}=\kappa
T_{\mu\nu}$.

Due to the form of our axially symmetric metric, the quantities
involved in the energy momentum tensor and the coefficients of the
perturbed conservation equations are functions of the coordinates
($R,z$) only. Let us consider a general perturbation $A^\mu_P$ of a
quantity $A^\mu$ in the form

\begin{equation}
A^\mu_P(t,R,\theta,z) = A^\mu(R,z) + \delta A^\mu(t,R,\theta,z),
\label{perturb}
\end{equation}

\noindent where $A^\mu(R,z)$ is the unperturbed quantity and $\delta
A^\mu(t,R,\theta,z)$ is the perturbation. Replacing (\ref{perturb}) for
each quantity in the energy momentum tensor (\ref{tmunu}) and
reordering terms, we obtain that

\begin{equation}
T^{\mu\nu}_P(t,R,\theta,z) = T^{\mu\nu}(R,z) + \delta
T^{\mu\nu}(t,R,\theta,z).
\end{equation}

Hereafter, we assume that the perturbed energy momentum tensor does not
modify the background metric found by solving the Einstein field
equation $G_{\mu\nu} = \kappa T_{\mu\nu}$. In other words, the
perturbation $\delta T^{\mu\nu}$ is treated as a test fluid. Not
considering the fluid as a test fluid is a second degree of
approximation to the stability problem in which the emission of
gravitational radiation is considered. With the perturbed energy
momentum tensor $T^{\mu\nu}_P$ and Einstein field equations, we obtain
that the perturbed energy momentum equations for the problem must obey

\begin{equation}
\delta T^{\mu\nu}_{;\nu} = 0. \label{tmunuperturb}
\end{equation}

\noindent These four equations can be found in Ref. \cite{uje:let} from
Eq. (8II) to Eq. (13II). Hereafter, the equations from Ref.
\cite{uje:let} will be denoted by II. In this manuscript we are
interested in linear perturbations of the energy momentum tensor. So,
in writing the explicit form of the four equations (\ref{tmunuperturb})
we discard terms of order greater or equal to ($\delta^2$). It is
explained in \cite{uje:let}, that if we want a consistent perturbation
model the tetrad must be perturbed. Moreover, demanding the orthonormal
condition for the perturbed tetrad we find several relations between
their components (14II). Because of the form of the perturbed energy
momentum equation for the axially symmetric metric (\ref{metric}), the
connection coefficients are only functions of the coordinates ($R,z$).
Therefore, all the coefficients on the perturbed conservation equations
depend only on ($R,z$) and we can construct a general perturbation of
the form

\begin{equation}
\delta A^\mu(t,R,\theta,z) = \delta A^\mu (R,z) e^{i(k_\theta
\theta- wt)}. \label{delta}
\end{equation}

Hereafter $\delta A^\mu \equiv \delta A^\mu (R,z)$. In this article, we
are interested in perturbations of the four velocity $U^\mu$ and the
thermodynamics variables ($\rho,p_R,p_\theta,p_z$). For that reason, we
see in (14II) that the $t$ components of the tetrad $\delta X^t$,
$\delta Y^t$ and $\delta Z^t$ must be perturbed. Moreover, the tetrad
perturbations $\delta X^\theta$, $\delta X^z$ and $\delta Y^z$ are not
related whatsoever to the four-velocity or the thermodynamic variables
perturbations and we can set them equal to zero. With these
assumptions, the perturbation (\ref{delta}) and conditions (14II), the
general perturbed energy momentum equations (8II)-(13II) can be cast
into the particular form

\begin{widetext}

$\mu=t$
\begin{eqnarray} \label{t}
&&\delta U^R_{,R} (\rho U^t + \xi_1 p_R X^R) + \delta U^z_{,z} (\rho U^t 
+ \xi_3 p_z Z^z) + \delta U^R [ {\rm F}(t,R,\rho U^t) + \xi_{1,R} p_R 
X^R + \xi_1 {\rm F}(t,R,p_R X^R)] \nonumber \\
&&+ \delta U^\theta [ i k_\theta (\rho U^t + \xi_2 p_\theta 
Y^\theta) ] + \delta U^z [ {\rm F}(t,z,\rho U^t) + \xi_{3,z} p_z Z^z + 
\xi_3 {\rm F}(t,z,p_z Z^z)] + \delta \rho (-i w U^t U^t) = 0,
\end{eqnarray}

$\mu=R$
\begin{eqnarray} \label{r}
&&\delta p_{R,R} (X^R X^R) + \delta U^R [-iw(\rho U^t +
\xi_1 p_R X^R)] + \delta \rho (U^t U^t \Gamma^R_{tt} ) \nonumber \\
&&+ \delta p_R {\rm G}(R,R,X^R X^R) + \delta p_\theta (Y^\theta 
Y^\theta \Gamma^R_{\theta\theta}) + \delta p_z (Z^z Z^z 
\Gamma^R_{zz}) = 0,
\end{eqnarray}

$\mu=\theta$ 
\begin{equation} \label{theta}
\delta U^\theta [-w (\rho U^t + \xi_2 p_\theta Y^\theta)] + 
\delta p_\theta (k_\theta Y^\theta Y^\theta) = 0,
\end{equation}

$\mu=z$
\begin{eqnarray} \label{z}
&&\delta p_{z,z} (Z^z Z^z) + \delta U^z [-iw(\rho U^t + \xi_3 p_z Z^z)] 
+ \delta \rho (U^t U^t \Gamma^z_{tt}) \nonumber \\
&&+ \delta p_R (X^R X^R \Gamma^z_{RR}) + \delta p_{\theta} (Y^\theta 
Y^\theta \Gamma^z_{\theta\theta}) + \delta p_z {\rm G}(z,z,Z^z Z^z) 
= 0.
\end{eqnarray} 
\noindent where
\begin{eqnarray}
&&{\rm F}(I,J,K) = K_{,J} + K (2 \Gamma^I_{IJ} + \Gamma^\alpha_{\alpha
J}), \\
&&{\rm G}(I,J,K) = K_{,J} + K (\Gamma^I_{IJ} + \Gamma^\alpha_{\alpha 
J}),
\end{eqnarray}
\noindent and $\Gamma^\alpha_{\beta\gamma}$ are the Christoffel symbols.
\end{widetext} 

Besides the four equations furnished by the energy momentum
conservation equations, $T^{\mu\nu}_{;\nu}=0$, there is another
important conservation equation, the equation of continuity, 

\begin{equation}
(n U^\mu)_{;\mu}=0, \label{continuity}
\end{equation}

\noindent where $n$ is the proper number density of particles. The
proper number density of particles $n$, and the total energy density
$\rho$ are related through the relation, 

\begin{equation}
\rho = n m_b +  \varepsilon, \label{varepsilon} 
\end{equation}

\noindent where $m_b$ is the constant mean baryon mass and
$\varepsilon$ the internal energy density. Multiplying Eq.
(\ref{varepsilon}) by $U^\mu$, performing the covariant derivative
($;\mu$) and using Eq. (\ref{continuity}), we obtain that

\begin{equation} 
(\rho U^\mu)_{;\mu} = (\varepsilon U^\mu)_{;\mu}. \label{rhovar}
\end{equation}

\noindent Now, from the relation $U_\nu T^{\mu\nu}_{;\mu}=0$ and the
energy momentum tensor (\ref{tmunu}), we obtain and expression for
$(\rho U^\mu)_{;\mu}$. Substituting this last expression into Eq.
(\ref{rhovar}) we finally arrive at

\begin{equation} 
(\varepsilon U^\mu)_{;\mu} = p_R X^\mu U_\nu X^\nu_{;\mu} + p_\theta
Y^\mu U_\nu Y^\nu_{;\mu} + p_z Z^\mu U_\nu Z^\nu_{;\mu}, \label{pdevar}
\end{equation}

\noindent which is a first order differential equation for
$\varepsilon$. Therefore, with $\varepsilon$ given by (\ref{pdevar})
the equation of continuity (\ref{continuity}) is satisfied. For this
reason, the continuity equation can be omitted for our stability
analysis because, in principle, we can always find a solution for
$\varepsilon$. Hereafter, the contribution of $n m_b$ and $\varepsilon$
to the total energy density are consider in $\rho$. In the case when
the internal energy density of the fluid is given, the equation of
continuity must be considered. In other words, the knowledge of the
thermodynamic properties of the disk is important for actual
applications. The thermodynamic properties of the system can be
obtained from observations or theoretically from the Fokker-Planck
equation, where we obtain the particle distribution function of the
disk. Solving the Fokker-Planck equation is not an easy task, but some
progress in Newtonian gravity have been done \citep{uje:let3}.

The four equations (\ref{t}-\ref{z}) contain seven independent
unknowns, say $\delta U^R, \delta U^\theta, \delta U^z, \delta \rho,
\delta p_R, \delta p_\theta, \delta p_z$. So, at this point, the number
of unknowns are greater than the number of equations and the system is
not self-consistent. This opens the possibility to perform different
kind of perturbations to make the system self-consistent. In Table
\ref{options} we present the classification for all the possible
perturbations that lead to a self-consistent system of equations. We
divided the table in three sections that depend on the number of
perturbed quantities, i.e. two, three and four quantities
perturbations. Equation (\ref{theta}) is important for this
classification because it only involve the two unknowns ($\delta
U^\theta,p_\theta$). Therefore, if one of them is, by construction,
equal to zero then the other unknown is also zero. There is a
possibility to avoid the other unknown to be equal zero, i.e. if the
unknown coefficient is equal zero. This situation can occur in
particular space-time points. We do not consider these cases. The
system of equations for the different combinations of the perturbed
quantities reduce to various differential equations, the form of the
particular differential equation is written in the column Differential
Equation. We denote by ODE\# and PDE\# an ordinary different equation
of order \# and a partial differential equation of order \#,
respectively. The differential equation marked with a question mark (?)
means that other four perturbations, that do not include the $\delta
U^\theta$ and $\delta p_\theta$ perturbations, are not self-consistent,
but they might be if we assume other imposed conditions for the involve
quantities, as for example $\delta p_R \equiv \delta p_z$. Moreover, if
we assume other imposed conditions for the different types of
perturbation shown in Table \ref{options}, the differential equation
classification may change.

We are interested in consistent physical perturbations. For that reason
we search for perturbations in which the velocity perturbation in a
certain direction leads to a pressure perturbation in the same
direction. For example, if we perturbed the $z$ component of the
velocity $\delta U^z$ then we must perturb $\delta p_z$ to have a
consistent physical problem. With the above criterion only four
perturbations are allowed, these are marked with (*) in Table
\ref{options}. These perturbations will be considered in our thick disk
model. Furthermore, we perform the perturbation $\delta U^R, \delta
p_R, \delta U^z, \delta p_z$ with the extra imposed condition $\delta
p_R \equiv \delta p_z$. This last perturbation are among the ones
denoted by question mark. In this particular case, the system of
equations reduced to a second order partial differential equation. 

\begin{table} 
\centering
\caption{Differential equation classification for all the possible
different types of perturbations. The perturbations marked with (*) are
the ones that are physical accepted. Other physical accepted
perturbations are marked with (?) but these perturbations require extra
imposed conditions.}

\label{options}
\begin{tabular}{|ccccc|} \hline
\multicolumn{4}{|c}{Perturbations} & \multicolumn{1}{c|}{Differential
Equation} \\ \hline
\multicolumn{5}{|c|}{{\bf Two Perturbations}} \\
& & $\delta p_R$ & $\delta p_z$ & PDE2 \\ 
\multicolumn{5}{|c|}{{\bf Three Perturbations}} \\
& $\delta U^R$ & $\delta p_R$ & $\delta U^z$ &  PDE2\\
& $\delta U^R$ & $\delta p_R$ & $\delta p_z$ &  ODE1 + PDE2\\
& $\delta U^R$ & $\delta p_R$ & $\delta \rho$ &  ODE2 *\\
& $\delta U^R$ & $\delta U^z$ & $\delta p_z$ &  PDE2\\
& $\delta U^R$ & $\delta U^z$ & $\delta \rho$ &  PDE1\\
& $\delta U^R$ & $\delta p_z$ & $\delta \rho$ &  PDE2\\
& $\delta p_R$ & $\delta U^z$ & $\delta p_z$ &  ODE1 + PDE2\\
& $\delta p_R$ & $\delta U^z$ & $\delta \rho$ &  PDE2\\
& $\delta U^z$ & $\delta p_z$ & $\delta \rho$ &  ODE2 *\\
\multicolumn{5}{|c|}{{\bf Four Perturbations}} \\
$\delta U^\theta$ & $\delta p_\theta$ & $\delta U^R$ & $\delta p_R$ &
 ODE2 *\\
$\delta U^\theta$ & $\delta p_\theta$ & $\delta U^R$ & $\delta U^z$ &
 PDE1\\
$\delta U^\theta$ & $\delta p_\theta$ & $\delta U^R$ & $\delta p_z$ &
 PDE2\\
$\delta U^\theta$ & $\delta p_\theta$ & $\delta U^R$ & $\delta \rho$ &
 ODE1\\
$\delta U^\theta$ & $\delta p_\theta$ & $\delta U^z$ & $\delta p_z$ &
 ODE2 *\\
$\delta U^\theta$ & $\delta p_\theta$ & $\delta U^z$ & $\delta p_R$ &
 PDE2\\
$\delta U^\theta$ & $\delta p_\theta$ & $\delta U^z$ & $\delta \rho$ &
 ODE1\\
$\delta U^\theta$ & $\delta p_\theta$ & $\delta p_R$ & $\delta \rho$ &
 ODE1\\
$\delta U^\theta$ & $\delta p_\theta$ & $\delta p_z$ & $\delta \rho$ &
 ODE1\\
\multicolumn{4}{|c}{Without $\delta U^\theta$, $\delta p_\theta$} &
\multicolumn{1}{c|}{?} \\
\hline
\end{tabular}
\end{table}

\section{Isotropic Schwarzschild Thick Disk}

The static isotropic Schwarzschild thick disk was found in
\cite{vog:let4} by applying the displace, cut, fill and replace method
to the axially symmetric metric (\ref{metric}) with the definitions

\begin{eqnarray}
&&\Psi_1 = \ln \left( \frac{2r - m}{2r + m} \right), \\
&&\Psi_2 = \Psi_3 = 2 \ln \left( 1 + \frac{m}{2r} \right),
\end{eqnarray}

\noindent where $r^2 = R^2 + (h+b)^2$, ($m>0$, $b>0$) are parameters,
and $h$ is a function of the coordinate $z$ given by

\begin{equation}
h(z) = \left\{
\begin{tabular}{ll}
$-z + C,$ & $z \le -a,$ \\
$A z^2 + B z^{2n+2}$, &  $-a \le z \le a,$ \\
$z + C,$ & $z \ge a,$
\end{tabular}
\right.
\end{equation}

\noindent with

\begin{eqnarray}
&&A = \frac{2n+1-ac}{4na}, \nonumber \\
&&B = \frac{ac-1}{4n(n+1)a^{2n+1}}, \nonumber \\
&&C = - \frac{a(2n+1+ac)}{4(n+1)}, \nonumber
\end{eqnarray}

\noindent where $n=1,2,\cdots$, and $c$ is the jump of the second
derivative on $z= \pm a$, see \cite{vog:let4}. The energy density and
the different pressures for this thick disk are

\begin{eqnarray} 
&&\rho = \frac{64m}{(m+2r)^5} [3(h+b)^2 (1- h^{\prime 2}) \nonumber \\
&&\hspace{3cm} + r^2 [h^{\prime 2}-1 + h^{\prime \prime}(h + b)]],
\label{rho} \\
&&p_R = p_\theta = \frac{32m^2}{(m+2r)^5(-m+2r)} [2(h+b)^2(1-h^{\prime
2}) \nonumber\\
&&\hspace{3cm}+ r^2 [h^{\prime 2} - 1 + h^{\prime \prime} (h + b)]],
\label{pr} \\
&&p_z = \frac{64 m^2 (1-h^{\prime 2}) (h+b)^2}{(m+2r)^5 (-m+2r)},
\label{pz}
\end{eqnarray}

\noindent where the prime ($\prime$) represents derivative with respect
to the axial coordinate $z$. We have set in Eqs.
(\ref{rho})-(\ref{pz}), without losing generalization, the range of the
$z$ coordinate between $-1 \le z \le 1$. These means that the parameter
$a$ of \cite{vog:let4} is equal to one. To satisfy the strong energy
condition we must have that $\epsilon = \rho + p_R +p_\theta +p_z \ge
0$, the weak energy condition requires $\rho \ge 0$ and the dominant
energy condition requires $|p_R/\rho| \le 1$, $|p_\theta/\rho| \le 1$
and $|p_z/\sigma| \le 1$. If we set the parameters ($m,b,n,c$)
explained in \cite{vog:let4} equal to ($m=1,b=2,n=1,c=0$), we obtain a
thick disk with all energy conditions satisfied. Also, the level curves
for this disk show that it is physically acceptable. For this set of
parameters, the function $h$ has the form $h = 3/4 z^2 -1/8 z^4$. We
remark that these are not the only parameters in which the level curves
are physically accepted. In the next section we applied the selected
perturbations of Sec. II to the isotropic Schwarzschild thick disk
mention above and studied its stability.

\section{Perturbations}

Before we start applying the different kinds of perturbations to the
isotropic Schwarzschild thick disk we must have some considerations.
Note that the exact thick disk metric considered is infinite in the
radial direction, but finite in the $z$ direction. In order to study
the stability we need a criterion to make a finite disk, i.e. we need a
cutoff in the radial coordinate. In Eqs. (\ref{rho}), (\ref{pr}) and
(\ref{pz}), we see that the thermodynamic quantities decrease rapidly
enough to define a cutoff radius. The cutoff radius $R_{cut}$ is set by
the following criterion: the energy density within the thick disk
formed by the cutoff radius is more than 90\% of the infinite thick
disk energy density. This leads to a cutoff radius of $R_{cut}=20$
units. Furthermore, the other 10\% of the energy density that is
distributed from outside the cutoff radius to infinity can be treated,
if necessary, as a perturbation in the outermost boundary condition.

\subsection{Perturbation with $\delta U^\theta$, $\delta p_\theta$,
$\delta U^R$, $\delta p_R$}

We start perturbing the four velocity in its components $\theta$ and
$R$, from the physical considerations mentioned in Sec. II we also
expect variations in the thermodynamic quantities $p_\theta$ and $p_R$.
The set of equations (\ref{t})-(\ref{z}) reduce to a second order
ordinary differential equation for the perturbation $\delta p_R$, say

\begin{equation}
F_A {\delta p_R}_{,RR} + F_B {\delta p_R}_{,R} + F_C \delta p_R = 0,
\label{edo2SIA}
\end{equation}

\noindent where $(F_A,F_B,F_C)$ are functions of ($R,z,w,k_\theta$),
see Appendix \ref{FSIA}.  For this particular case we have for the
perturbation, $\delta p_\theta = - \delta p_R$. 

Note that in Eq. (\ref{edo2SIA}) the coordinate $z$ only enters like a
parameter. Moreover, the equation for $\delta p_R$ is independent of
the parameter $w$, but $w$ needs to be different from zero to reach
that form. The second order equation (\ref{edo2SIA}) is solved
numerically with two boundary conditions, one at $R \approx 0$ and the
other at the cutoff radius. At $R \approx 0$ we set the perturbation
$\delta p_R$ to be $\approx$ 10\% of the unperturbed pressure $p_R$
(\ref{pr}). In the outer radius of the disk we set $\delta
{p_R}|_{R=R_{cut}} = 0$ because we want our perturbation to vanish when
approaching the edge of the disk, and in that way, to be in accordance
with the linear perturbation applied. We say that our perturbations are
valid if their values are lower, or of the same order of magnitude,
than the 10\% values of the unperturbed quantities.

\begin{figure*}
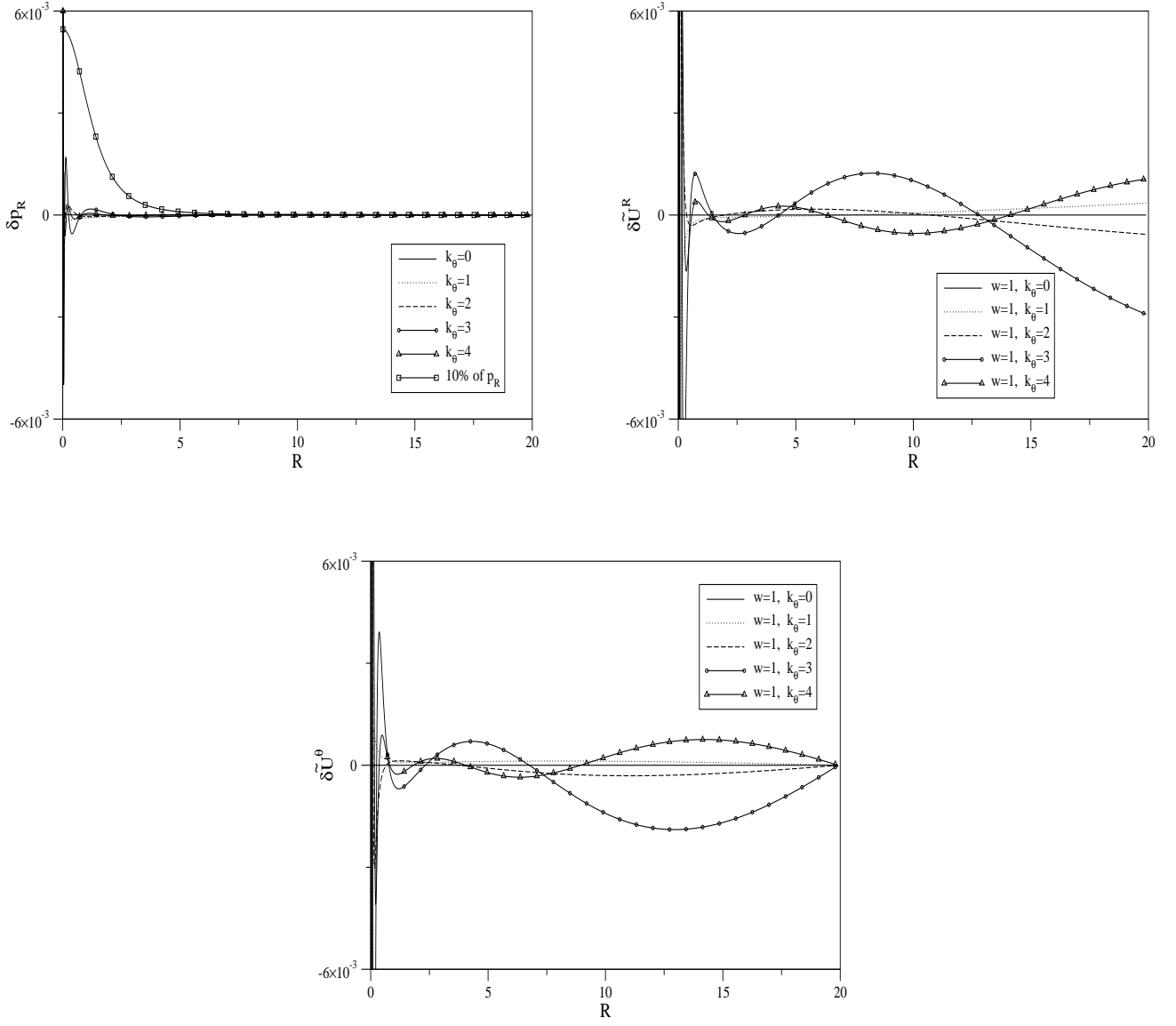
 \vspace{0.7cm}
\epsfig{width=8cm, height=7cm, file=fig1a.eps} \hspace{1cm} 
\epsfig{width=8cm, height=7cm, file=fig1b.eps}

\vspace{1.2cm} 

\epsfig{width=8cm, height=7cm, file=fig1c.eps}

\vspace{0.2cm}

\caption{Profiles of the amplitude perturbations modes for the radial
pressure, radial physical velocity and azimuthal physical velocity for
the case when $z=0$, $w=1$ and $k_\theta=0,1,2,3,4$. In the pressure
perturbation graph the 10\% profile value of the radial pressure is
depicted to ensure that the perturbations are in accordance with the
stability criterion imposed. The value of the velocity perturbation
profiles are well below the particles escape velocity.} 

\label{figSIA}
\end{figure*}

In Fig. \ref{figSIA}, we present the amplitude profile of the radial
pressure perturbation, the physical radial velocity perturbation
$\tilde{\delta U^R} = \sqrt{g_{RR}} \delta U^R$ and the physical
azimuthal velocity perturbation $\tilde{\delta U^\theta} =
\sqrt{g_{\theta \theta}} \delta U^\theta$ in the plane $z=0$. In Fig.
\ref{figSIA} we see that the perturbation $\delta p_R$ decreases
rapidly with $R$ and has oscillatory behavior. At first sight, the
perturbation $\delta p_R$ appears to be stable for all $R$, but in
order to make a complete analysis we have to compare at each radius the
values of the different perturbations with the values of the radial
pressure. For this purpose, we included in the same graph a profile of
the 10\% value of $p_R$. We see that the perturbations of $\delta p_R$
for different values of $k_\theta$ are always lower or, at least, of
the same order of magnitude when compared to these 10\% values.

Our four velocity $U^\mu$ (\ref{tetrad}) has only components in the
temporal part, so we do not have values of $U^R$ and $U^\theta$ to make
comparisons with the perturbed values $\tilde{\delta U^R}$ and
$\tilde{\delta U^\theta}$ of Fig \ref{figSIA}. For that reason we
compared, in first approximation, the amplitude profiles of these
perturbations with the value of the escape velocity in the Newtonian
limit. In the Newtonian limit of General Relativity, we have the well
known relation $g_{00} = \eta_{00} + 2 \Phi$. So, the Newtonian escape
velocity $V_e = \sqrt{2 |\Phi|}$ can be written as

\begin{equation} 
V_e = \left( 1 - \frac{(1-\frac{m}{2r})^2}{(1+\frac{m}{2r})^2}
\right)^{1/2}.
\end{equation}

\begin{figure*}
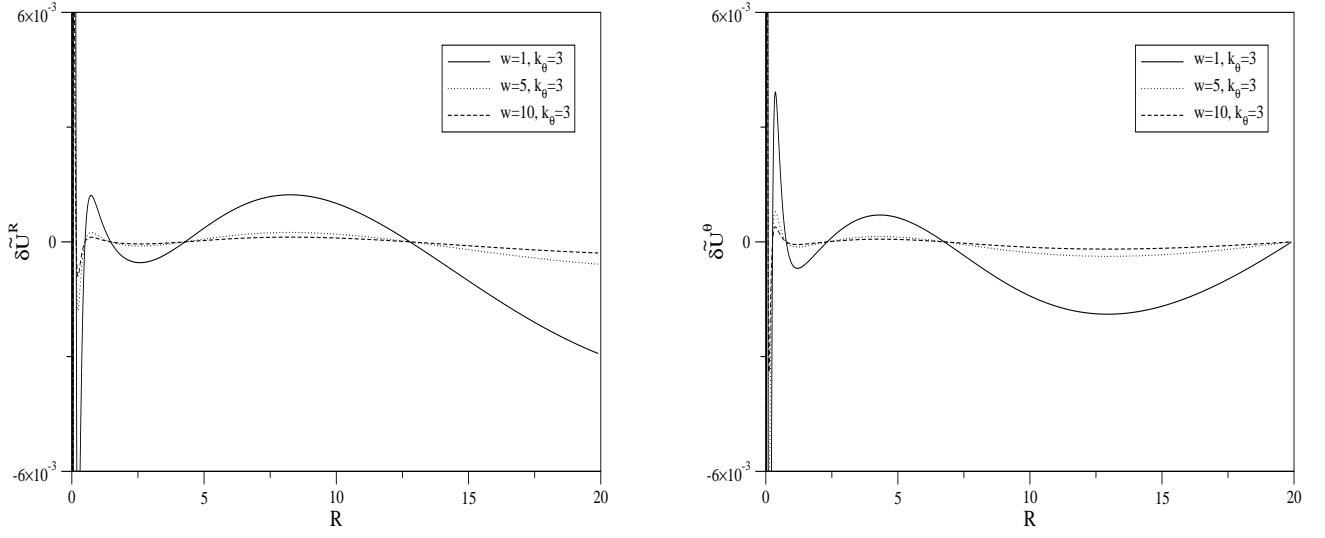
 \vspace{0.7cm}
\epsfig{width=8cm, height=7cm, file=fig2a.eps} \hspace{1cm} 
\epsfig{width=8cm, height=7cm, file=fig2b.eps}

\vspace{0.2cm}

\caption{Profiles of the amplitude perturbations modes for the radial
physical velocity and the azimuthal velocity for the case when $z=0$,
$k_\theta=3$ and $w=1,5,10$. We note that when we increase the value
of the frequency $w$ the modes are more stable.} 

\label{figSIAW}
\end{figure*}

For the parameters considered, we have that at $(R=0,z=0)$ and
$(R=20,z=0)$ the escape velocity is $V_e=0.8$ and $V_e \approx 0.31$,
respectively. The escape velocity values have a small variation with
$z$. So, with this criterion, we may say that the perturbations
$\tilde{\delta U^R}$ and $\tilde{\delta U^\theta}$ are stable because
their values are always well below the escape velocity value. Recall
that the perturbation $\delta p_R$ does not depend on the parameter
$w$, but the perturbations $\tilde{\delta U^R}$ and $\tilde{\delta
U^\theta}$ do. In Fig. \ref{figSIAW} we show the numerical solution of
$\tilde{\delta U^R}$ and $\tilde{\delta U^\theta}$ for different values
of the frequency $w$ with the same wavenumber $k_\theta$. We see that
when we increase the value of the parameter $w$ the perturbations
become more stable. This behavior is the same for other values of
$k_\theta$.

In this subsection we set the value of the parameter $z=0$. We
performed the same analysis for different values of the parameter
$-1<z<1$, and we found that the perturbations show the same qualitative
behavior. Therefore, we can said that the isotropic thick disk is
stable under perturbations of the form presented in this subsection.
Nevertheless, if we treat the 10\% of the energy density as a
perturbation in the outermost radius of the disk by setting $\delta p_R
|_{R=R_{cut}} = \epsilon$, where $\epsilon < 10\%$ of
$p_R|_{R=R_{cut}}$ the qualitative behavior of the mode profiles are
the same. From these results, we can say that the general linear
perturbation applied is stable, and for that reason, the perturbations
do not form more complex structures like rings, bars or halos.
Moreover, if we set the frequency $w \rightarrow iw$ we obtain the same
equation for the perturbation $\delta p_R$, say (\ref{edo2SIA}). In
this case, the real part of the general perturbation (\ref{delta})
diverge with time and the perturbation is not stable. These last
considerations can be applied to every perturbation in the following
subsections.

\subsection{Perturbation with $\delta U^\theta$, $\delta p_\theta$,
$\delta U^z$, $\delta p_z$}

In this subsection we perturb the four velocity in its components
$\theta$ and $z$, and we expect variations in the thermodynamic
quantities $p_\theta$ and $p_z$. The set of equations
(\ref{t})-(\ref{z}) reduce to a second order ordinary differential
equation for the perturbation $\delta p_z$ given by

\begin{equation}
F_A {\delta p_z}_{,zz} + F_B {\delta p_z}_{,z} + F_C \delta p_z = 0,
\label{edo2SIB}
\end{equation}

\noindent where $(F_A,F_B,F_C)$ are functions of ($R,z,w,k_\theta$),
see Appendix \ref{FSIB}. Note that in Eq. (\ref{edo2SIB}) the
coordinate $R$ only enters like a parameter. Like the previous case,
Eq. (\ref{edo2SIB}) is independent of the parameter $w$, but in order
to reach that form we must have $w$ different from zero. The second
order equation (\ref{edo2SIB}) is solved numerically with two boundary
conditions, one in $z = 0$ and the other in $z =1$. At $z=0$ we set the
perturbation $\delta p_z$ to be $\approx 10\%$ of the unperturbed
pressure $p_z$ (\ref{pz}). In the outer plane of the disk we set
$\delta p_z |_{z=1} = 0$ because we want our perturbation to vanish
when approaching the edge of the disk, and in that way, to be in
accordance with the linear perturbation applied.

\begin{figure*}
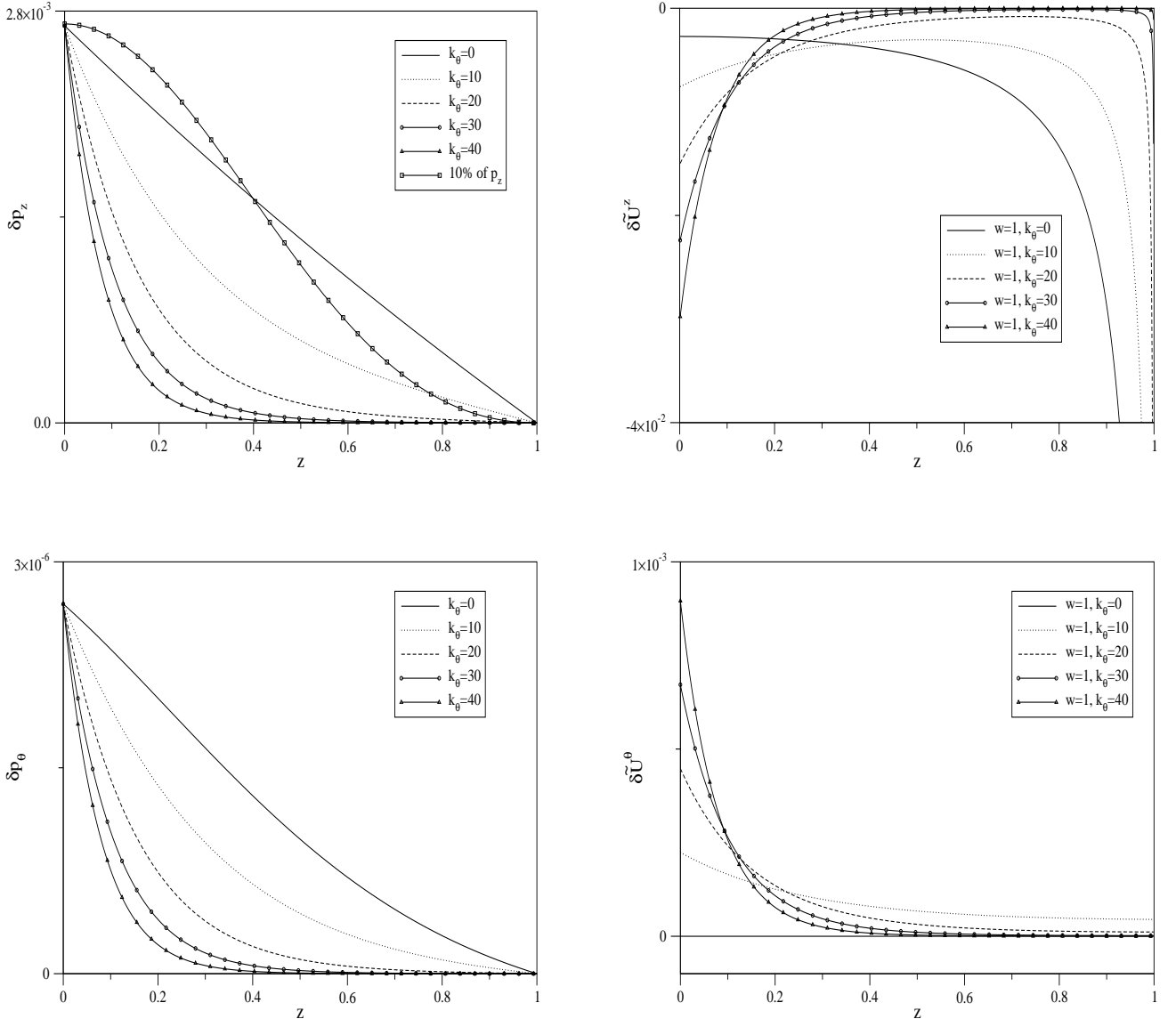
 \vspace{0.7cm}
\epsfig{width=8cm, height=7cm, file=fig3a.eps} \hspace{1cm} 
\epsfig{width=8cm, height=7cm, file=fig3b.eps}

\vspace{1.2cm} 

\epsfig{width=8cm, height=7cm, file=fig3c.eps} \hspace{1cm}
\epsfig{width=8cm, height=7cm, file=fig3d.eps}

\vspace{0.2cm}

\caption{Profiles of the amplitude perturbations modes for the axial
pressure, azimuthal pressure, axial physical velocity and azimuthal
physical velocity for the case when $R=0.1$, $w=1$ and
$k_\theta=0,10,20,30,40$. In the pressure perturbation graph the 10\%
profile value of the axial pressure is depicted to ensure that the
perturbations are in accordance with the stability criterion imposed.
In the azimuthal pressure perturbation graph the 10\% profile was
omitted because the values are three order of magnitudes higher. The
values of the azimuthal velocity perturbation profiles are well below
the particles escape velocity and the modes are stable. For the axial
velocity perturbation profiles we see that the geometric properties of
the disk create a kind of wall that impedes the particles for going
out.} 

\label{figSIB}
\end{figure*}

In Fig \ref{figSIB}, we present the amplitude profiles of the axial
pressure perturbation, the azimuthal pressure perturbation, the
physical axial velocity $\tilde{\delta U^z}=\sqrt{g_{zz}} \delta U^z$
and the physical azimuthal velocity for the value of the parameter
$R=0.1$. For comparison reasons, we included in the graph of the axial
pressure perturbation the amplitude profile that corresponds to 10\% of
the value of $p_z$. We see in this graph that the modes profiles of the
axial pressure are below, or of the same order of magnitude, when
compared to the 10\% profile. In this example the modes with wavenumber
$k_\theta=0$ and $k_\theta=10$ have within the disk regions in which
the profile values exceed the 10\% profile. This perturbations are near
the validity criterion used for the perturbations. In the axial
velocity perturbations graph we see that the profiles suffer a sudden
variation near the edge of the disk $z=1$. The axial velocity
perturbation tends to expel the particles out from the disk. In our
case, the particles are not allow to escape from the disk due to the
fact that our geometrical disk is only valid for $-1 \le z \le 1$. For
that reason, the particles feel a kind of a wall that impedes them from
going out. It is clear from the axial velocity perturbation profiles
that the two nodes $(w=1, k_\theta=0)$ and $(w=1, k_\theta=10)$ are not
stable, while the rest of the nodes are stable if we discard the region
that is near the edge of the disk at $z=1$. Note that the starting
point of the sudden variation of the axial velocity perturbation
coincides with the radius in which the axial pressure perturbation
exceeds the 10\% value profile of $p_z$. Also, in Fig. \ref{figSIB}, we
see that the azimuthal pressure perturbations and the azimuthal
velocity perturbations for different modes are stable. In the case of
the azimuthal pressure, we have not included the 10\% profile because
is three order of magnitude higher than the perturbation value. Note
that the amplitude values of azimuthal velocity perturbation are below
the escape velocity considered for the disk. 

\begin{figure*} \vspace{0.7cm}
\epsfig{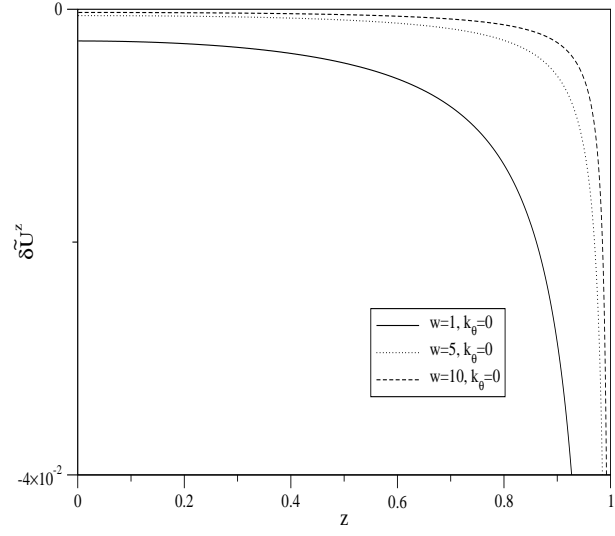} 

\vspace{0.2cm}

\caption{Profiles of the amplitude perturbations modes for the axial
physical velocity for the case when $R=0.1$, $k_\theta=0$ and
$w=1,5,10$. We note that if we increase the value of the $w$ parameter,
the region in which the mode is stable increases. For higher values of
$w$ the modes are stable if we discard the region near $z=1$.} 

\label{figSIBW}
\end{figure*}

The perturbation $\delta p_z$ does not depend on the parameter $w$, but
the perturbations $\tilde{\delta U^z}$ and $\tilde{\delta U^\theta}$ do
depend. In Fig. \ref{figSIBW}, we show the profile for the mode
$k_\theta = 0$ with different values of the frequency $w$ for the
perturbation $\tilde{\delta U^z}$. We see that when we increase the
value of the parameter $w=1$, the velocity perturbations are more
stable (if we neglect the region near the edge of the disk at $z=1$).
This is due to the fact that the frequency $w$ only enters the equation
for $\delta U^z$ like a scale factor.

We have performed the same above analysis for different values of the
parameter $R$, and we found that the qualitative behavior is the same.
With these results we can say, if we neglect the region near the edge
of the disk at $z=1$, that the isotropic thick disk is always stable
for higher values of the parameter $w$ and has some stable modes for
lower values of $w$. Moreover, for lower values of the frequency $w$,
some modes of our general first order perturbation are not stable. In
these cases, more complex structures may be form, but in order to study
them, we need a higher order perturbation formalism.

\subsection{Perturbation with $\delta U^R$, $\delta p_R$, $\delta
\rho$}

In this subsection, we perturb the radial component of the four
velocity, the radial pressure and the energy density of the fluid. The
set of equations (\ref{t})-(\ref{z}) reduce to a second order ordinary
differential equation for the perturbation $\delta p_R$ of the form
(\ref{edo2SIA}). The form of the functions $(F_A,F_B,F_C)$ are given in
Appendix \ref{FSIC}. In this case, the coordinate $z$ only enters like
a parameter. Due to the fact that we are not considering perturbations
in the azimuthal axis, the coefficients of the second order ordinary
diffetential equation do not depend on the wavenumber $k_\theta$. This
second order equation is solved numerically with the same boundary
conditions described in Sec. IV-A.

\begin{figure*}
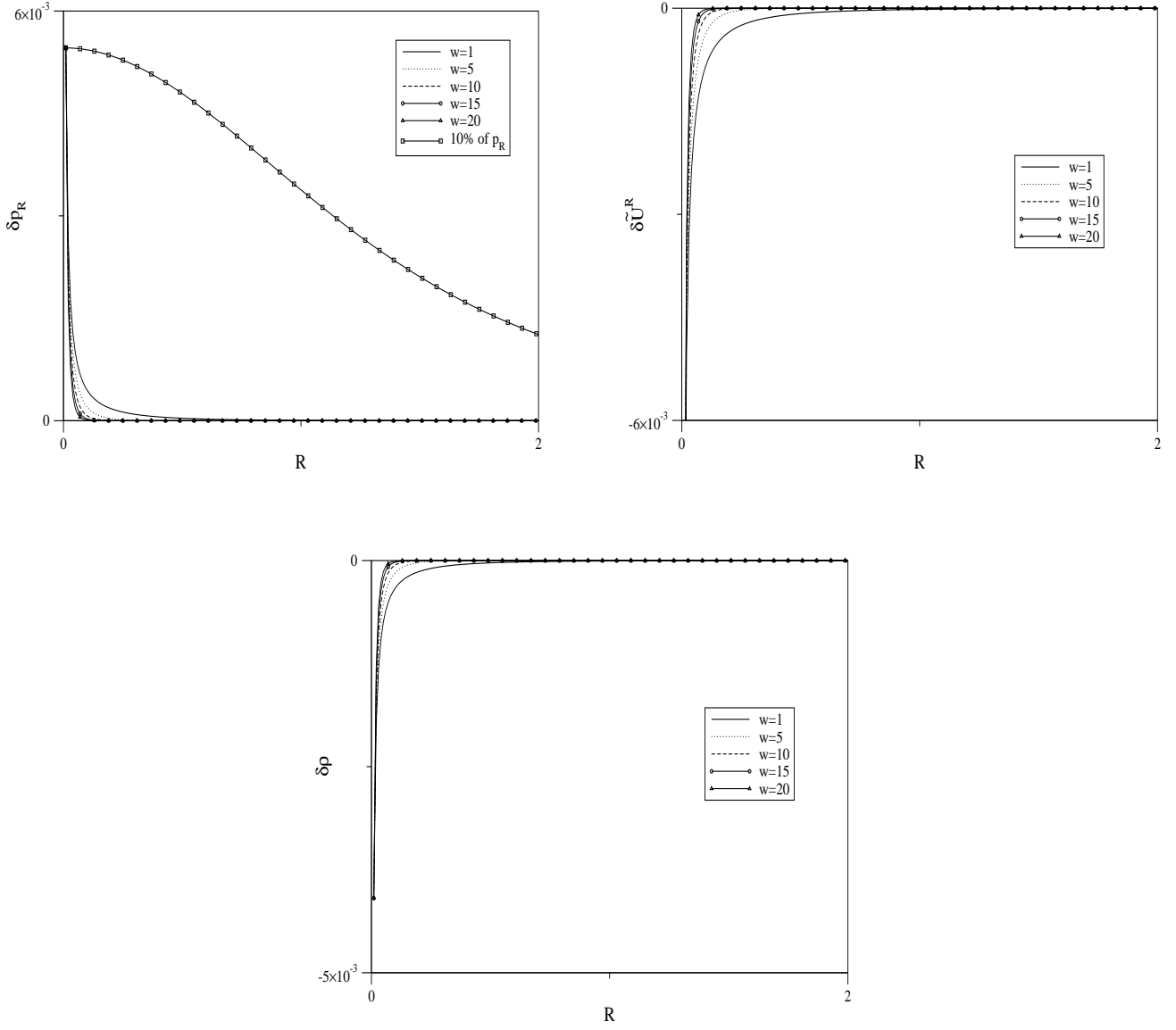
 \vspace{0.7cm}
\epsfig{width=8cm, height=7cm, file=fig5a.eps} \hspace{1cm} 
\epsfig{width=8cm, height=7cm, file=fig5b.eps}

\vspace{1.2cm} 

\epsfig{width=8cm, height=7cm, file=fig5c.eps}

\vspace{0.2cm}

\caption{Profiles of the amplitude perturbations modes for the radial
pressure, radial physical velocity and energy density for the case when
$z=0$ and $w=1,5,10,15,20$. As in Fig. \ref{figSIA}, we have depicted
the 10\% profile value of the radial pressure in the pressure
perturbation graph. We see from these graphs that the pressure
perturbation values are well below the 10\% profile. In the energy
density perturbation graph the 10\% profile of the energy density was
omitted because the values are two order of magnitudes higher. Also,
the values of the velocity perturbations profiles are always lower than
the escape velocity. Therefore, the modes are highly stable.}

\label{figSIC}
\end{figure*}

In Fig. \ref{figSIC} we present the amplitude profiles for different
perturbation modes of the axial pressure, physical axial velocity and
energy density for the value of the parameter $z=0$. We see in the
graphs, that the perturbation profiles decrease rapidly in few units of
$R$. Note that the graphs in Fig. \ref{figSIC} are plotted in the range
[0,2], but the radius of the disk has been set in 20 units. We do not
plot the 10\% profile of the energy density because is at least two
order of magnitude higher than the values depicted. Also, the values of
the radial perturbation modes profiles are well below the escape
velocity. We can say that all these modes are highly stable. We
performed the above analysis for different values of $z$ and we found
that the quantities involve have the same qualitative behavior. From
these results, we can say that the general linear perturbation applied
is highly stable, and for that reason, the perturbations do not form
more complex structures like rings, bars or halos.

\subsection{Perturbation with $\delta U^z$, $\delta p_z$, $\delta
\rho$}

In this subsection we perturb the $z$ component of the four velocity,
the $z$ component of the pressure and the energy density of the fluid.
The set of equations (\ref{t})-(\ref{z}) reduce to a second order
ordinary differential equation for the perturbation $\delta p_z$ of the
form (\ref{edo2SIB}). The functions $(F_A,F_B,F_C)$ are given in
Appendix \ref{FSID}. Note that, like in Sec. IV-B, the coordinate $R$
enters like a parameter. In this case, we are not considering azimuthal
perturbations and therefore the quantities involve do not depend on the
parameter $k_\theta$. The second order equation is solved following the
procedure of Sec. IV-B.

\begin{figure*}
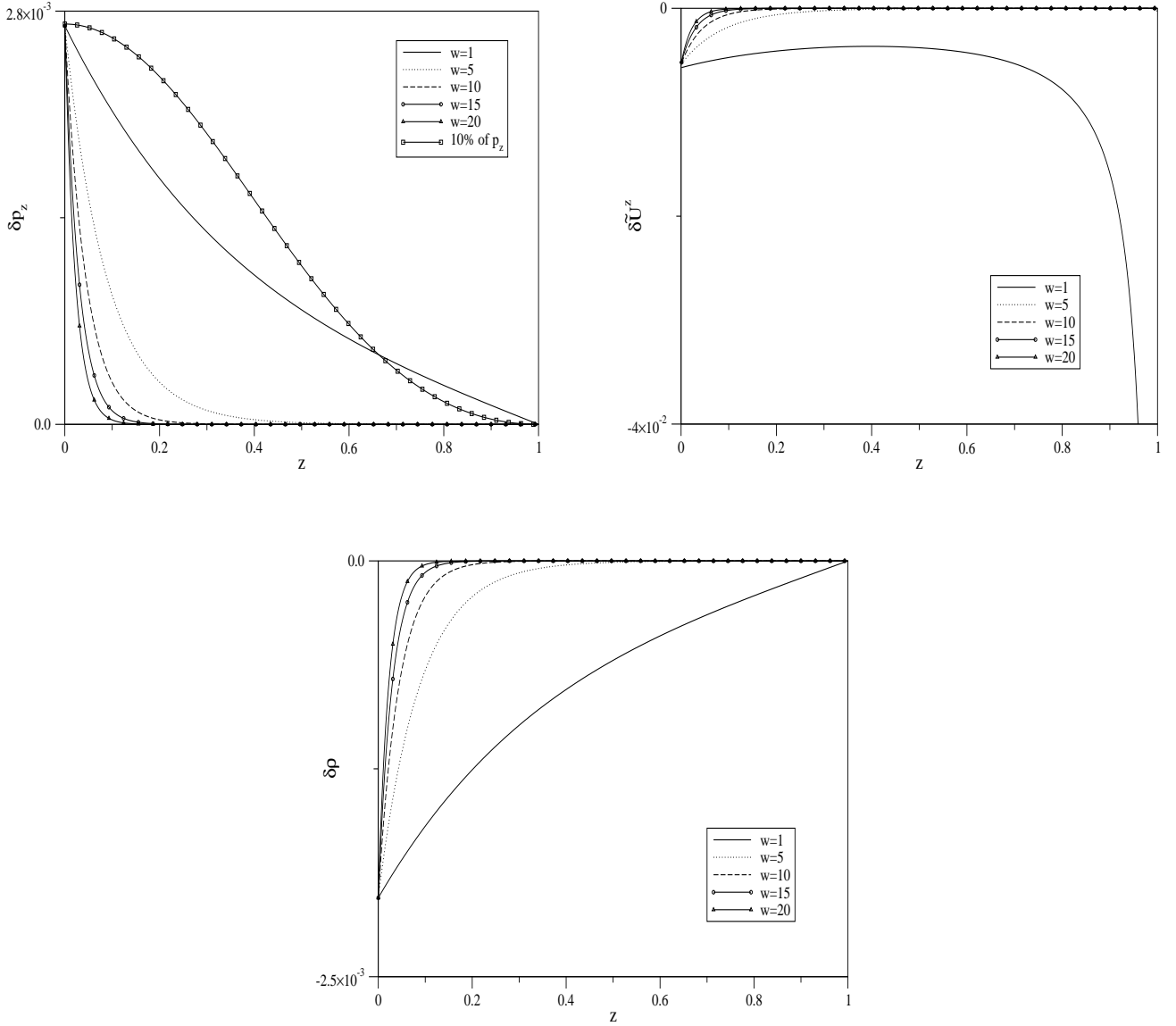
 \vspace{0.7cm}
\epsfig{width=8cm, height=7cm, file=fig6a.eps} \hspace{1cm} 
\epsfig{width=8cm, height=7cm, file=fig6b.eps}

\vspace{1.2cm} 

\epsfig{width=8cm, height=7cm, file=fig6c.eps}

\vspace{0.2cm}

\caption{Profiles of the amplitude perturbations modes for the axial
pressure, axial physical velocity and energy density for the case when
$R=0.1$ and $w=1,5,10,15,20$. To ensure that the perturbations are in
accordance with the stability criterion imposed, we have depicted in
the pressure perturbation graph the 10\% profile of the axial pressure.
In the energy density perturbation graph the 10\% profile of the energy
density was omitted because the values are two order of magnitudes
higher. In the axial velocity perturbation profiles the mode with $w=1$
is not stable. Note that the starting radius of the abrupt increase in
the axial velocity perturbation amplitude coincide with the point in
which the pressure perturbation becomes greater than the 10\% profile.}

\label{figSID}
\end{figure*}

In Fig. \ref{figSID} we present the amplitude profiles of the axial
pressure perturbation, the axial velocity perturbation and the energy
density perturbation for the value of the parameter $R=0.1$. We see
that the the pressure perturbation modes are always of the some order
of magnitude or lower when compared to the 10\% profile. Similar to the
Sec. IV-B case, the velocity perturbation amplitude profiles increase
abruptly near the edge of the disk at $z=1$. For higher values of the
parameter $w$ the modes are stable. Even for the most stable mode there
is a abruptly increase in the amplitude profile at the edge of the
disk. This change in the behavior occurs very near $z=1$ and can not be
seen in the axial velocity perturbation graph. In this case we can say
that the mode with $w=1$ is not stable. If we compare the axial
velocity perturbation graph of  Fig. \ref{figSID} with Fig.
\ref{figSIBW}, we note that the perturbation modes considered in this
subsection have larger regions of stability. The modes that correspond
to the energy density perturbations are all stable. Note that we have
not plot the 10\% profile of the energy density because is at least two
order of magnitude higher than the values depicted. We performed the
above analysis for different values of the parameter $R$ and we found
that the quantities involve have the same qualitative behavior. Similar
to Sec. IV-B, some modes corresponding to lower values of the frequency
$w$ are not stable. As we note in Fig. \ref{figSID}, this is related to
the fact that our first order perturbation is not valid in some
regions. In these cases, more complex structures may be form, but in
order to study them, we need a higher order perturbation formalism.

\subsection{Perturbation with $\delta U^R$, $\delta p_R$,
$\delta U^z$, $\delta p_z$ and $\delta p_R \equiv \delta p_z$}

In this subsection we perturb the radial component of the four
velocity, the axial component of the four velocity, the radial pressure
and the axial pressure. This kind of perturbation belongs to the ones
marked with a question mark in Table \ref{options}. As we said in Sec.
II, we need an extra condition to make the set of equations
self-consistent. In this case, we set $\delta p_R \equiv \delta p_z
\equiv \delta p$. Therefore, the set of equations (\ref{t})-(\ref{z})
reduce to a second order partial differential equation for the pressure
perturbation $\delta p$, say

\begin{equation}
F_A \delta p_{,RR} + F_B \delta p_{,R} + F_C \delta p_{,zz} + F_D \delta
p_{,z} + F_E \delta p = 0, \label{pde2SIE}
\end{equation}

\noindent where ($F_A,F_B,F_C,F_D,F_E$) are functions of ($R,z,w$), see
Appendix \ref{FSIE}. The partial differential equation (\ref{pde2SIE})
is solved numerically with four boundary conditions, at $z \approx -1$,
$z \approx 1$, $R \approx 0$ and $R = 20$. The boundary condition at
$z$ can not be set at 1 because some quantities diverge. As we said in
Sec. IV-B, this is due to the geometrical restrictions that has the
isotropic Schwarzschild thick disk. They are different ways in which we
can set the boundary conditions in order to simulate various kinds of
pressure perturbations. Here, we treat only the case when we have a
pressure perturbation at $R \approx 0$ and along the $z$ axis, i.e.
some kind of a rod perturbation. We set the value of the rod pressure
perturbation to be 10\% of the axial pressure. For the rest of the
boundary conditions we set the values equal to zero because we want the
perturbation to vanish when approaching the edge of the disk. We chose
the 10\% of the value of the axial pressure instead of the radial
pressure because it has the lowest value near $R \approx 0$. In that
way, the perturbation values are also below the 10\% values of the
radial pressure and the general linear perturbation is valid.

\begin{figure*} \vspace{0.7cm}
\epsfig{width=8cm, height=7cm, file=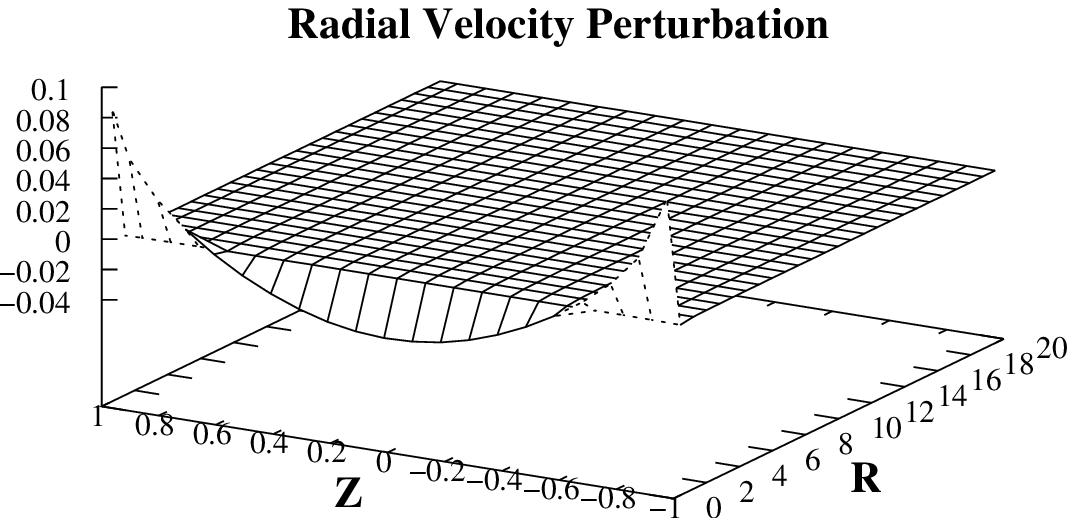} \hspace{1cm} 
\epsfig{width=8cm, height=7cm, file=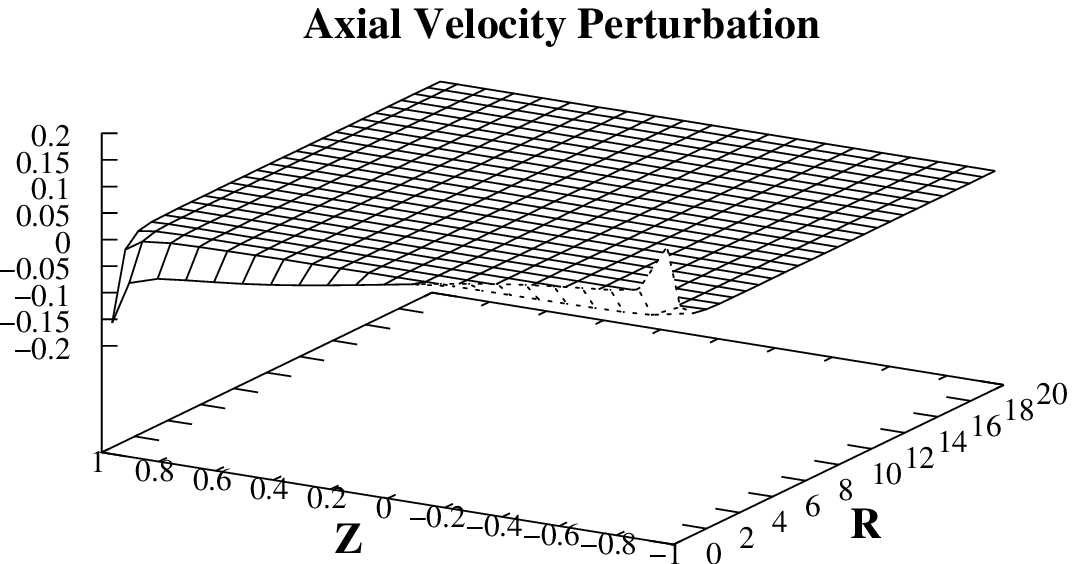}

\vspace{1.2cm} 

\epsfig{width=8cm, height=7cm, file=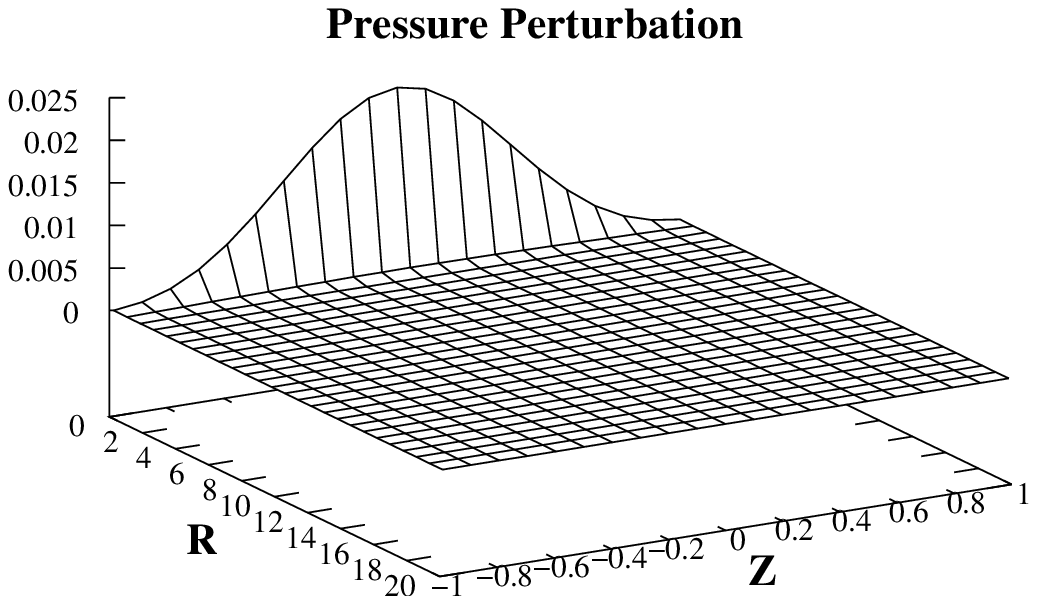}

\vspace{0.2cm}

\caption{Profiles of the amplitude perturbations modes for the
pressure, radial physical velocity and axial physical velocity for the
case when $w=1$. We see that the perturbation rapidly goes to values
near zero when we depart from the center of the disk, i.e. the
perturbations are stable. The parameter $w$ enters as a scale factor in
the differential equations and do not alter the qualitative properties
of the perturbations.}

\label{figSIE}
\end{figure*}

In Fig. \ref{figSIE}, we present the perturbation amplitude for the
pressure, the physical radial velocity and the physical axial velocity
for $w=1$. We see in the pressure perturbation graph that the
perturbation rapidly decays to values near zero when we move out from
the center of the disk. In the velocity perturbations profiles we can
see an interesting phenomenon. Note that in the lower domain of the
disk [-1,0) the axial velocity perturbation is positive and in the
upper domain (0,1] the axial velocity perturbation is negative. This
means that due to the linear perturbation the disk tries to collapse to
the plane $z=0$. Now, if we look to the radial velocity perturbation
graph, we note that the upper and lower parts depart from the center of
the disk due to the positive radial perturbation. The center of the
disk tries to condensate. So, with these considerations, we may say
that the disk tends to form some kind of a ring around the center of
the disk. This phenomenon appears near the center of the disk in a few
units of $R$. The parameter $w$ enters only like a scale factor in the
differential equations quantities studied.

\section{Conclusions}

In this manuscript we obtain the basic perturbation equations for
studying the stability of thick disks. These equations were obtained
when we applied a general first order perturbation to the conservation
equations of motion. The great number of unknowns make the set of
equations not self-consistent. In order to make the system of equations
self-consistent we must chose a combination of the perturbed
quantities. This opens the possibility for various types of
combinations. In this manuscript we made a classification of these
possibilities and studied the physical accepted perturbations. We can
say, that the stability analysis we performed is more complete than the
stability analysis of particle motion along geodesics because we take
into account the collective behavior of the particles. However, this
analysis can be said to be incomplete because the energy momentum
perturbation tensor of the fluid is treated as a test fluid and does
not alter the background metric. This is a second degree of
approximation to the stability problem in which the emission of
gravitational radiation is considered. 

The stability analyses made to the isotropic Schwarzschild thick disk,
show that this disk is stable when we applied radial and azimuthal
perturbations. In the case of axial perturbations, the nodes are not
stable near the edge of the disk. The lack of stability is due to the
form of the geometric thick disk considered \cite{vog:let4}, i.e. it is
only valid in a fixed axial interval $-a \le z \le a$, where $a$ is a
constant (in our case $a$=1). If we discard the region near the axial
edge we can say, in general, that for higher values of the parameters
$w$ and $k_\theta$ the axial modes are also stable. For lower values of
$w$, some axial perturbations are not stable. This is related to the
fact that our first order perturbation is not valid in some regions of
the considered quantities. In these cases, more complex structures may
be form, but in order to study them, we need a higher order
perturbation formalism.

\section*{ACKNOWLEDGMENTS}

M.U. and P.S.L. thanks FAPESP for financial support; P.S.L. also
thanks CNPq.

\appendix

\section{Coefficients of perturbation type A} \label{FSIA}

The general form of the functions ($F_A,F_B,F_C$) appearing in the
second order ordinary differential equation (\ref{edo2SIA}) are given
by

\begin{eqnarray} 
&&F_A=A_1 \alpha_1, \hspace{0.5cm} F_B=A_1
{\alpha_1}_{,R} + A_1 \alpha_2 + A_3 \alpha_1, \nonumber \\ 
&&F_C=A_1 {\alpha_2}_{,R} + A_3 \alpha_2 + A_4 \alpha_3, \label{A1} 
\end{eqnarray}

\noindent where $\alpha_1$, $\alpha_2$ and $\alpha_3$ are

\begin{eqnarray}
&&\alpha_1 = -\frac{B_1}{B_2}, \hspace{0.5cm} 
\alpha_2 = \frac{B_5 D_4 - B_4 D_5}{B_2 D_5}, \nonumber \\
&&\alpha_3 = \frac{C_2 D_4}{C_1 D_5}. \label{A2}
\end{eqnarray}

\noindent In Eqs. (\ref{A1}) and (\ref{A2}), we denote the coefficients
of Eq. (\ref{t}) by $A_i$, the coefficient of Eq. (\ref{r}) by $B_i$,
the coefficient of Eq. (\ref{theta}) by $C_i$, the coefficient of Eq.
(\ref{z}) by $D_i$, e.g., the first term in (\ref{t}) has the
coefficient $A_1$ multiplied by the factor $\delta U^R_{,R}$, the
second term has the coefficient $A_2$ multiplied by the factor $\delta
U^R$, etc. The explicit form of the above equations is obtained
replacing the fluid variables ($\rho,p_R,p_\theta,p_z$) of the
isotropic Schwarzschild thick disk.

\section{Coefficients of perturbation type B} \label{FSIB}

The general form of the functions ($F_A,F_B,F_C$) appearing in the
second order ordinary differential equation (\ref{edo2SIB}) are given
by

\begin{eqnarray} 
&&F_A=A_2 \alpha_1, \hspace{0.5cm} F_B=A_2
{\alpha_1}_{,z} + A_2 \alpha_2 + A_5 \alpha_1, \nonumber \\ 
&&F_C=A_2 {\alpha_2}_{,z} + A_4 \alpha_3 + A_5 \alpha_2,
\end{eqnarray}

\noindent where $\alpha_1$, $\alpha_2$ and $\alpha_3$ are

\begin{eqnarray}
&&\alpha_1 = -\frac{D_1}{D_2}, \hspace{0.5cm} 
\alpha_2 = \frac{B_6 D_5 - B_5 D_6}{B_5 D_2}, \nonumber \\
&&\alpha_3 = \frac{C_2 B_6}{C_1 B_5},
\end{eqnarray}

\noindent where the meaning of the notation for the coefficients
($A_i,B_i,C_i,D_i$) are explained in Appendix \ref{FSIA}.

\section{Coefficients of perturbation type C} \label{FSIC}

The general form of the functions ($F_A,F_B,F_C$) are given by

\begin{eqnarray} 
&&F_A=A_1 \alpha_1, \hspace{0.5cm} F_B=A_1
{\alpha_1}_{,R} + A_1 \alpha_2 + A_3 \alpha_1, \nonumber \\ 
&&F_C=A_1 {\alpha_2}_{,R} + A_3 \alpha_2 + A_6 \alpha_3,
\end{eqnarray}

\noindent where $\alpha_1$, $\alpha_2$ and $\alpha_3$ are

\begin{eqnarray}
&&\alpha_1 = -\frac{B_1}{B_2}, \hspace{0.5cm} 
\alpha_2 = \frac{B_3 D_4 - B_4 D_3}{B_2 D_3}, \nonumber \\
&&\alpha_3 = - \frac{D_4}{D_3},
\end{eqnarray}

\noindent where the meaning of the notation for the coefficients
($A_i,B_i,D_i$) are explained in Appendix \ref{FSIA}.

\section{Coefficients of perturbation type D} \label{FSID}

The general form of the functions ($F_A,F_B,F_C$) are given by

\begin{eqnarray} 
&&F_A=A_2 \alpha_1, \hspace{0.5cm} F_B=A_2
{\alpha_1}_{,z} + A_2 \alpha_2 + A_5 \alpha_1, \nonumber \\ 
&&F_C=A_2 {\alpha_2}_{,z} + A_5 \alpha_2 + A_6 \alpha_3,
\end{eqnarray}

\noindent where $\alpha_1$, $\alpha_2$ and $\alpha_3$ are

\begin{eqnarray}
&&\alpha_1 = -\frac{D_1}{D_2}, \hspace{0.5cm} 
\alpha_2 = \frac{B_6 D_3 - B_3 D_6}{B_3 D_2}, \nonumber \\
&&\alpha_3 = - \frac{B_6}{B_3},
\end{eqnarray}

\noindent where the meaning of the notation for the coefficients
($A_i,B_i,D_i$) are explained in Appendix \ref{FSIA}.

\section{Coefficients of perturbation type E} \label{FSIE}

The general form of the functions ($F_A,F_B,F_C,F_D,F_E$) appearing in
the partial second order differential equation (\ref{pde2SIE}) are
given by

\begin{eqnarray}  
&&F_A=A_1 \alpha_1, \hspace{0.5cm} F_B=A_1 {\alpha_1}_{,R} + A_1
\alpha_2 + A_3 \alpha_1, \nonumber \\  
&&F_C=A_2 \alpha_3, \hspace{0.5cm} F_D=A_2 {\alpha_3}_{,z} + A_2
\alpha_4 + A_5 \alpha_3, \nonumber \\ 
&&F_E = A_1 {\alpha_2}_{,R} + A_2 {\alpha_4}_{,z} + A_3 \alpha_2 + A_5
\alpha_4, \label{E1} 
\end{eqnarray}

\noindent where $\alpha_1$, $\alpha_2$, $\alpha_3$ and $\alpha_4$ are

\begin{eqnarray}
&&\alpha_1 = -\frac{B_1}{B_2}, \hspace{0.5cm} \alpha_2 = -\frac{B_4 +
B_6}{B_2}, \nonumber \\
&&\alpha_3 = -\frac{D_1}{D_2}, \hspace{0.5cm} \alpha_4 = -\frac{D_4 +
D_6}{D_2}. \label{E2}
\end{eqnarray}

\noindent where the meaning of the notation for the coefficients
($A_i,B_i,D_i$) are explained in Appendix \ref{FSIA}.



\begin{thebibliography}{99}
\bibitem{uje:let2} In the Appendix section of reference \cite{uje:let}
are some typographical errors. In Eq. (A1) we must perform the
transformation $\alpha_1 \leftrightarrow \alpha_2$. In Eq. (B3) there
is a $(p_{\varphi,r}/\sigma_{,r})$ factor missing.
\end{thebibliography}
\end{document}